\documentclass[letterpaper, 10 pt, conference]{ieeeconf}  % Comment this line out
\IEEEoverridecommandlockouts                              % This command is only
\usepackage[utf8]{inputenc}
\usepackage[T1]{fontenc}

\usepackage{amsmath} % assumes amsmath package installed
\usepackage{amssymb}  % assumes amsmath package installed

\usepackage{subcaption}
\usepackage{multicol} 
\usepackage{stackrel}
\usepackage{graphicx}
\usepackage{listings}
\usepackage{csquotes}
\usepackage[hidelinks]{hyperref}
\usepackage{makeidx}
\usepackage{cancel}
\usepackage{xcolor}
\usepackage{braket}
\usepackage{bm} 
\title{\LARGE \bf
Telegraph Processes with Extrinsic Fluctuations: Burst Dynamics and an Application to Domestic Cat Activity}

\author{Manuel Eduardo Hernández-García$^{2}$ and Monica S. López-Castaños$^{1}$% <-this % stops a space
\thanks{$^{1}$Especialidad en Biotecnología Aplicada, Laboratorio de Biología Molecular y Glicosilación 
en Cáncer, Facultad de Ciencias Químicas, Benemérita Universidad Autónoma de Puebla, Heroica Puebla de Zaragoza 72570, México.
Email:mlopezcastanos@gmail.com}
\thanks{$^{2}$Facultad de Ciencias Físico Matemáticas, Benemérita Universidad Autónoma de Puebla, Heroica Puebla de Zaragoza 72570, México.
Email: manuel.hernandezgarcia@viep.com.mx }%
}

\begin{document}

\maketitle
\thispagestyle{empty}
\pagestyle{empty}

%%%%%%%%%%%%%%%%%%%%%%%%%%%%%%%%%%%%%%%%%%%%%%%%%%%%%%%%%%%%%%%%%%%%%%%%%%%%%%%%
\begin{abstract}
\begin{abstract}
Animal activity often consists of intermittent bursts separated by prolonged periods of inactivity. Here, we describe this behavior using a two-state telegraph process with extrinsically fluctuating transition rates. We derived analytical expressions for the stationary behavior of the system and characterized how stationary fluctuations in the effective transition rates modified the activity and residence time statistics. In particular, we show that whereas fixed transition rates lead to exponential residence-time distributions, stationary rate fluctuations generate effective Lomax distributions with heavier tails. We also found that fluctuations can either increase or decrease the mean occupancy of the active state, and that when the fluctuations in the two transition rates are equal, their effect on the stationary occupancy becomes indistinguishable from the case without fluctuations. We assessed the parameter inference using synthetic data and applied the framework to observations of a domestic cat. For synthetic data, accounting for extrinsic fluctuations substantially improved the recovery of the kinetic parameters used to generate the data. In the experimental data, the inferred variability of the inactive-to-active transition rate was substantially larger than that of the active-to-inactive rate. These results establish a tractable framework for studying stochastic switching systems in which extrinsic variability reshapes burst and residence-time statistics.
\end{abstract}
\end{abstract}

\tableofcontents 

%%%%%%%%%%%%%%%%%%%%%%%%%%%%%%%%%%%%%%%%%%%%%%%%%%%%%%%%%%%%%%%%%%%%%%%%%%%%%%%%
\section{Introduction}

Biological systems are inherently subject to both intrinsic and extrinsic fluctuations. Intrinsic fluctuations arise from the stochastic nature of (molecular) interactions \cite{Gar}, whereas extrinsic fluctuations reflect the variability in kinetic parameters induced by environmental fluctuations \cite{Extrinsic, Hilfinger}. Rather than being mere perturbations, these sources of variability fundamentally shape the dynamics and observable behaviors of biological systems \cite{Singh, Elo}. 

Animal behavioral activity is characterized by temporal fluctuations and intermittent transitions between the active and inactive states. In domestic cats (\textit{Felis catus}), locomotor behavior exhibits a pronounced temporal organization, with periods of activity interspersed with relatively long periods of rest or inactivity \cite{AssessingCircadian}. Continuous monitoring studies have shown that cats display daily rhythmic patterns of locomotor activity, often with bimodal peaks associated with the dawn and dusk periods \cite{ForinWiart, Parker}. More recent studies using accelerometry have explicitly characterized feline behavior in terms of active and inactive states, providing a quantitative description of the temporal structure of these transitions \cite{Garcia, AssessingCircadian}. These observations indicate that activity is not a stationary process but rather a dynamically fluctuating behavior influenced by both internal and environmental factors.

The activity of domestic cats is sensitive to external conditions. Environmental enrichment, human interactions, feeding schedules, and the predictability of the surrounding environment can modify the occurrence and duration of behavioral activity \cite{ Stella}. In particular, unpredictable environmental events have been associated with changes in activity levels and other behavioral responses, suggesting that external perturbations can alter the temporal organization of the active--inactive cycle \cite{ Stella}. At the same time, substantial inter-individual differences have been observed in behavioral responses, indicating that intrinsic characteristics may also contribute to fluctuations and transitions between behavioral states \cite{CopingStyles}.

From a mathematical perspective, these observations motivate the description of feline behavior as a stochastic dynamic process in which animals intermittently switch between active and inactive states. Rather than treating activity as a deterministic periodic signal, this framework allows both intrinsic fluctuations and extrinsic perturbations to be incorporated into dynamics. Therefore, the central question is as follows: How do stationary fluctuations in effective transition rates shape the temporal dynamics of activity and inactivity?

In this study, we addressed this question by introducing a generalized two-state model based on the telegraph process \cite{Wang} describing the transitions between inactive and active states in domestic cats. Although the standard telegraph model has been widely used to capture switching dynamics, such as gene expression bursting \cite{Holehouse, Jiao, Zhang} and financial processes \cite{Ratanov}, it typically assumes constant transition rates. Here, we extend this framework by incorporating extrinsically fluctuating transition rates and derive closed-form analytical expressions for the stationary behavior of the system.

We show that extrinsic variability induces nontrivial modifications to the statistical properties of the system, leading to qualitative changes in the probability distributions of activity. In particular, we demonstrate that parameter fluctuations can generate effective heavy-tailed behaviors \cite{Ham} and significantly reshape burst statistics beyond the standard telegraph predictions. These theoretical results were validated using both synthetic and experimental data from observations of a domestic cat, establishing a minimal yet powerful framework for quantifying how environmental fluctuations reshape behavioral. We chose the likelihood method \cite{Chen, Myung}, for simplicity, although other more robust methods exist to infer parameters \cite{Sukys, Kim2013, Tan}. 

The remainder of this paper is organized as follows. In Section \ref{sec:2}, we present a model to describe the active state of domestic cats. In Section \ref{sec:3}, we derive the probability distribution without parameter fluctuations. In Section \ref{sec:4}, we analyze the system by incorporating extrinsic fluctuations. In Section \ref{sec:5}, we present how to obtain experimental data from observing a cat and how to infer the parameters of the model. Finally, in Section \ref{sec:6}, we present the conclusions of the study. 

\section{Model}

\begin{figure}[tbh!]
\begin{subfigure}{\linewidth}
\centering
\includegraphics[width=0.69\columnwidth]{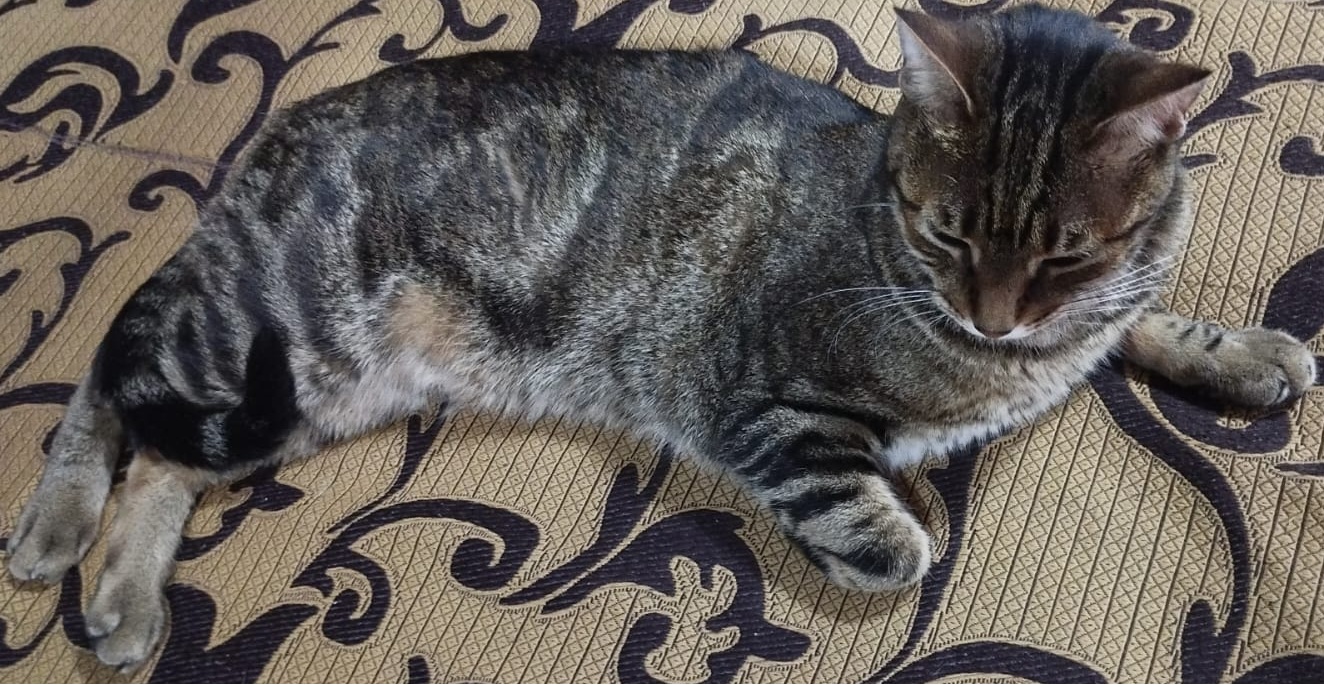}
\includegraphics[width=0.25\columnwidth]{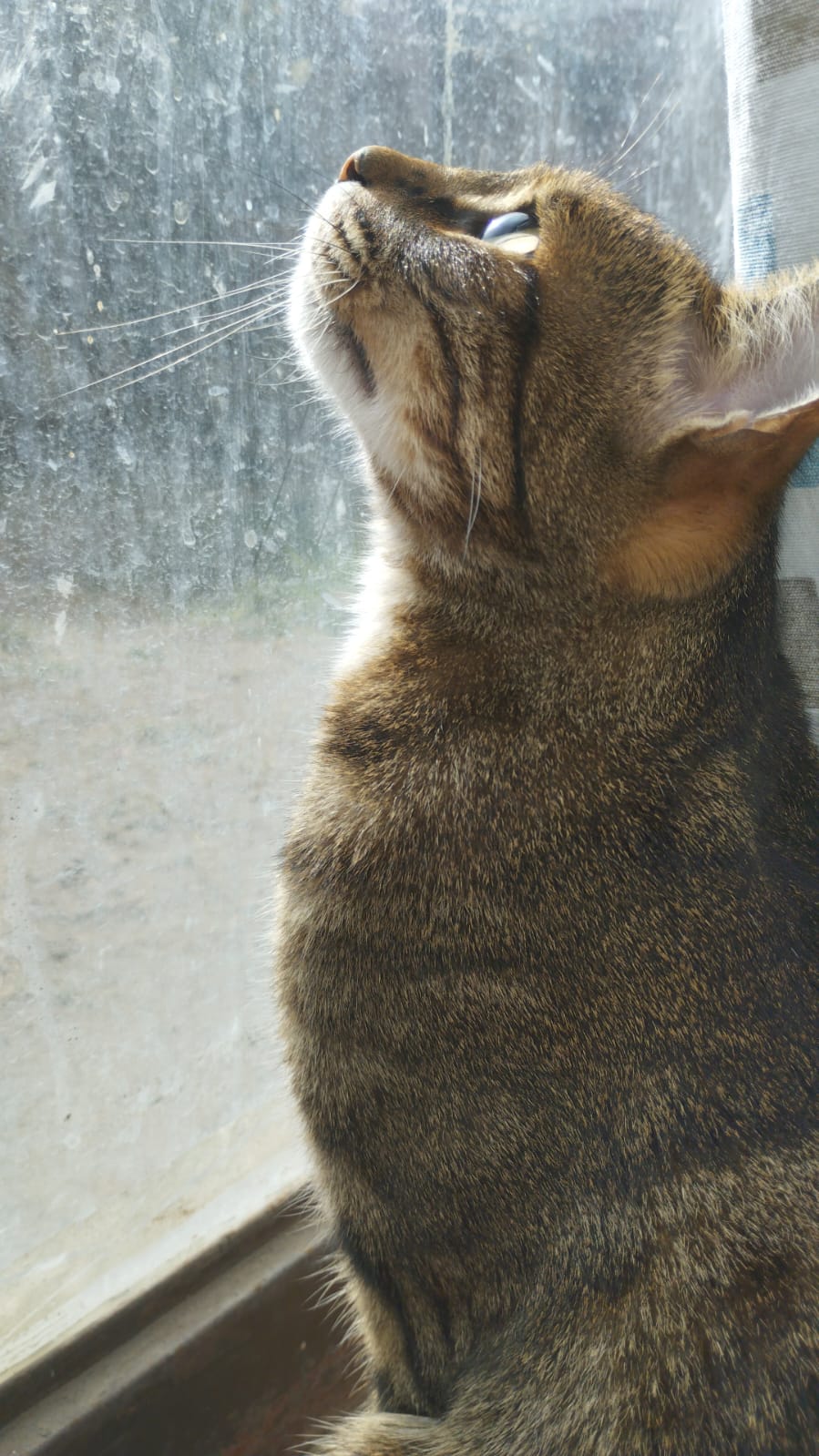}
\caption{}
\end{subfigure}
\begin{subfigure}{\linewidth}
\centering
\includegraphics[width=0.7\columnwidth]{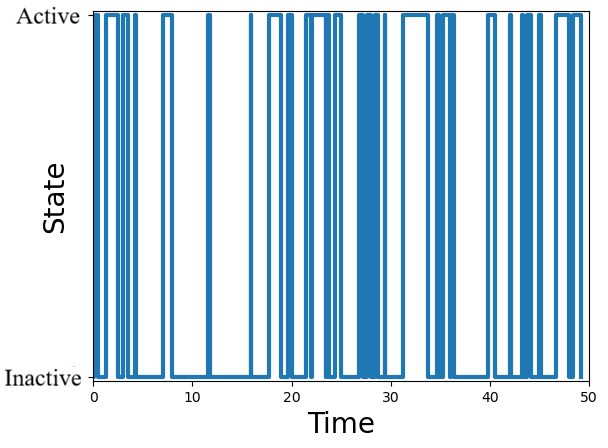}
\caption{}
\end{subfigure}\par\medskip
\caption{\textbf{Telegraph processes.} (a) This panel illustrates inactive and active states in domestic cats. 
(b) The telegraph process is plotted to show how the system transitions between active and inactive states. The parameters used were $K_{I}=1$ and $K_{A}=2$.  }
\label{fig:1}
\end{figure}

Domestic cats remain inactive for relatively long periods, interspersed with intermittent bursts of activity \cite{Parker, AssessingCircadian}. To describe this behavior, we considered two distinct behavioral states (Fig. \ref{fig:1}(a)). We defined the inactive state (I) as periods during which the cat remained stationary or asleep, whereas the active state (A) encompassed sustained observable activities, such as walking, grooming, or observing the surroundings. We deliberately coarse-grained the behavioral repertoire into two states, active and inactive, because our objective was to characterize the stochastic dynamics of behavioral switching rather than to distinguish specific behavioral categories.

To describe this behavior, we can formulate a model inspired by the telegraph process \cite{Holehouse, Jiao, Zhang}, which can be expressed as follows:
\begin{table}[h]
    \centering
    \begin{tabular}{|c|c|}
        $ C_I \stackbin{K_{I}}{\longrightarrow} C_A$ & $C_A  \stackbin{K_{A}}{\longrightarrow} C_I$ 
    \end{tabular}
\end{table}

A transition from an inactive to an active state occurs when the cat gets up and initiates sustained activity. Conversely, short pauses during an active episode are not considered a transition to inactivity unless the cat returns to a stationary or resting state. Therefore, the two transitions in the model represent the onset and termination of the behavioral activity. The kinetic parameter $K_I$ denotes the transition rate from inactivity to activity, or activation rate, whereas $K_A$ denotes the transition rate from activity to inactivity, or deactivation rate, respectively. These rates determined the characteristic time spent by the cat in each behavioral state. The resulting switching dynamics are analogous to those commonly used in stochastic models of gene expression \cite{Holehouse, Jiao, Zhang, Tunna}.

Let $T_{K_I}$ and $T_{K_A}$ denote the residence times in the inactive and active states, respectively. For fixed transition rates, these quantities characterize the durations of the individual inactive and active episodes.

We assume that during each behavioral episode, the transition rates $K_I$ and $K_A$ remain approximately constant. However, the effective rates may differ between successive episodes because of fluctuations in the variables that modulate behavioral changes. In this framework, extrinsic fluctuations are represented as episode-to-episode variability in transition rates rather than continuous changes occurring within a single episode.

\section{Distributions without Extrinsic Fluctuations} \label{sec:3}

First, we present the case in the absence of extrinsic fluctuations and calculate the corresponding distributions.  The master equation describing the processes is \cite{Iyer},

\begin{subequations} \label{1}
{\footnotesize
    \begin{align}
        \partial_t P(C_I,t)=& - K_{I} P(C_I,t) + K_{A} P(C_A,t), \\
        \partial_t P(C_A,t)= 
        &K_{I} P(C_I,t) - K_{A} P(C_A,t),
    \end{align}}
\end{subequations}
the simulation of these processes is shown in Fig. \ref{fig:1}(b), where the transitions between states can be clearly observed. Writing Eq. \eqref{1} in matrix form, introducing $\mathbf{P}(state)= [P(C_I,t), P(C_A,t)]$, $state=(C_I, C_A)$ and
\begin{align}
    \mathbf{W}= \left(\begin{matrix}
        -K_{I} & K_{A} \\
        K_{I} & -K_{A}
    \end{matrix} \right),
\end{align}
the solution of Eq. \eqref{1} is \cite{Gar}
{\footnotesize
\begin{align}
    \mathbf{P}(state,t)= \mathbf{P}(state,0) + \frac{1-e^{-2K t}}{2K} \mathbf{W}\mathbf{P}(state,0),
\end{align}}
where $K= \frac{K_{I}+ K_{A}}{2}$ is the mean of the kinetic parameters and $\mathbf{P}(state,0)$ is the initial state. Because the distribution approaches the stationary distribution exponentially, we used the stationary approximation. This assumption is also justified by considering adult cats, whose behavioral patterns are often considered stable over time and, therefore, approximated as stationary. The stationary solution approaches a stationary distribution $\mathbf{P}(state,t \to \infty)= \mathbf{P_s}(state)$
\begin{align}
    \mathbf{P_s}(state)=  \frac{1}{K_{I}+K_{A}} \left(\begin{matrix}
        K_{A} \\
        K_{I} 
    \end{matrix} \right)   . \label{3}
\end{align}
Note that $P_s(C_I)+ P_s(C_A)=1$. In the following, we use the stationary distribution given by Eq. \eqref{3}.

\begin{figure}[h]
\centering
\includegraphics[width=0.45\columnwidth]{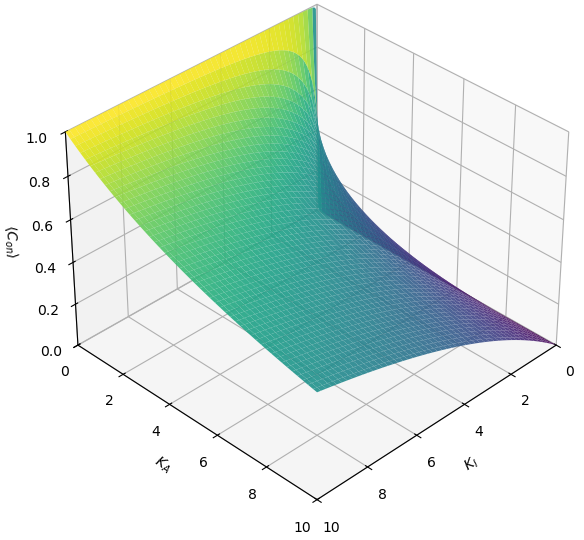}
\includegraphics[width=0.45\columnwidth]{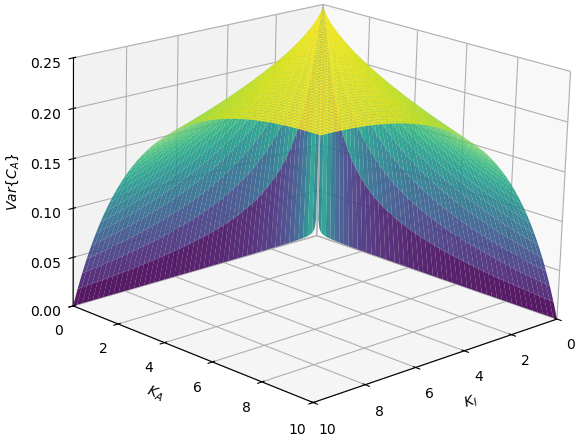}
\caption{\textbf{Mean and Variance of Active State.} In the left panel, we plot the mean of the active state for different values of $K_{I}$ and $K_{A}$. We observe that the mean lies between $0$ and $1$, and it reaches small values only when $K_{I}$ is small. In the right panel, we plot the variance of the active state, where the maximum variance occurs when $K_{I}=K_{A}$.} 

\label{fig:2}

\end{figure}

The mean and variance of the active state are
\begin{subequations}
    \begin{align}
        \braket{C_A}_s=& \frac{K_{I}}{K_{I}+ K_{A}}, \\
        Var\{C_A\}_s=& \braket{C_A}_s  \left( 1- \braket{C_A}_s \right).
    \end{align}
\end{subequations}
In Fig. \ref{fig:2}, we plot the mean and variance of the active state. The results show that the mean value of the active state lies between $0$ and $1$, and the variance lies between $0$ and $0.25$. 

When $K_{I}>K_{A}$, transitions from the inactive to the active state occur more frequently; therefore, the system remains in the active state for a longer time. However, the time spent in the active state is small. As $K_{A}$ increases, these bursts become shorter and less persistent, leading to a decrease in the mean active state. 

In contrast, when $K_{I}<K_{A}$, transitions into the active state are rare, resulting in a small number of bursts. This behavior can be interpreted as a period during which the cat remains inactive for extended periods before returning to an active state, possibly reflecting the time required to recover energy after episodes of intense activity.

However, the variance reaches its maximum when $K_{I}=K_{A}$. This corresponds to the regime in which the probability of a burst event is approximately $0.5$, leading to the largest fluctuations in the active state because both states occur with equal probability. When $K_{I}\ll 1$ and $K_{A}\ll 1$, the variance is low. Interestingly, the shape of the variance surface resembled the ears of a cat.

\section{Impact of Extrinsic Fluctuations on Distributions} \label{sec:4}

The transition rate $K_{I}$ represents the effective rate of activation, which may be influenced by external conditions, such as random object movements, ambient sounds, or vibrations. In contrast, $K_{A}$ represents the rate at which the cat leaves the active state and may be influenced by internal factors such as stimulus saturation, fatigue, or the need for sleep. These factors are not necessarily independent, and both transition rates may be affected by a combination of external and internal conditions.

Because environmental conditions are inherently variable and physiological states may also fluctuate \cite{Stella,CopingStyles}, it is natural to allow the effective transition rates $K_{I}$ and $K_{A}$ to also fluctuate \cite{Extrinsic}. These fluctuations provide an additional source of variability in the switching dynamics.

We did not explicitly model the temporal dynamics of the variables responsible for these fluctuations. Instead, we assumed that they had reached a stationary regime and characterized their effects using the stationary distributions of the effective transition rates. This assumption is appropriate when considering adult cats whose behavioral patterns remain relatively stable over time and when the animal is observed under environmental conditions that do not undergo drastic changes, such as remaining in the same household environment for extended periods.

Let $\theta=\{ K_{I}, K_{A} \}$ be the set of kinetic parameters of the system. Since transition rates are positive quantities, we represent their stationary fluctuations using a gamma distribution \cite{Gorin} ($\rho \in \theta$)
\begin{align}
        P_{\rho}(\rho)=& \frac{\left( \frac{\rho}{\sigma_{\rho}} \right)^{\left( \frac{\underline \rho}{\sigma_{\rho}}-1\right)} e^{ - \frac{\rho}{\sigma_{\rho}}  }}{\sigma_{\rho}\Gamma\left( \frac{\underline \rho}{\sigma_{\rho}}\right)}  , \label{13}
\end{align}
$\Gamma(x)$ is the gamma function. This distribution depends on a set of parameters $\Theta_{\rho}=(\sigma_{\rho}, \underline \rho)$ (with $(\sigma_{\rho}<1)$), which we refer to as hyperparameters. To better understand the role of these, we compute the mean and variance of the kinetic parameter from the distribution, 
\begin{subequations}
    \begin{align}
        \braket{\rho}=& \underline \rho, \\
        Var\{\rho\}=&   \sigma_{\rho} \underline \rho.
    \end{align}
\end{subequations}
Then, $\underline \rho$ is the mean, and $\sigma_{\rho}$ adjusts the size of the variance, which is proportional to $\underline \rho$. Therefore, \(\sigma_\rho\) controls the magnitude of the extrinsic fluctuations in the corresponding transition rate.

The transitions between the active and inactive states, considering that each kinetic parameter has fluctuations, where $K_{I}$ and $K_{A}$ follow a gamma distribution. It has the following full distribution, 

{\footnotesize
\begin{align}
    \mathbf{P}_{full}(state, K_{I},K_{A})=& \mathbf{P}_s(state|K_{I},K_{A})  P_{\underline K_{I}}(K_{I}) P_{\underline K_{A}}(K_{A}).
\end{align}}
Where $\mathbf{P}_s(state|K_{I},K_{A})$ is the result obtained in Eq. \eqref{3}. To determine the distributions of the active and the inactive state over all possible values of the kinetic parameters, we evaluate 

{\small
\begin{align} \label{14}
    \mathbf{P}_{pf}(state)= \int_{0}^{\infty} \int_{0}^{\infty} dK_{I} dK_{A} \mathbf{P}_{full}(state, K_{I},K_{A}), 
\end{align}}
and we obtain
{\footnotesize
\begin{subequations}
    \begin{align}
    \mathbf{P}_{pf}(state)= &\left(\begin{matrix}
        1-a \\
        a 
    \end{matrix} \right), \\
    a=&  \frac{\alpha_{I}}{\alpha_{A}+\alpha_{I}} {}_{2}F_{1}\left(1, \alpha_{A}; \alpha_{I} + \alpha_{A}+1, 1- \frac{\sigma_{I}}{\sigma_{A}} \right), 
\end{align}
\end{subequations}}
with $\alpha_{\rho}= \frac{\underline \rho}{\sigma_{\rho}}$ and ${}_{2}F_{1}$ the ordinary hypergeometric function, then the mean and the variance in the active state are 
\begin{subequations}
{\small
\begin{align}
\braket{C_A}_{pf}=& a, \\
Var\{C_A\}_{pf}=&\braket{C_A}_{pf} \left( 1- \braket{C_A}_{pf} \right).
\end{align}}
\end{subequations}
In this case, the mean of the active state also depends on terms that involve the variances $\sigma_{\rho}$, which modify its value compared to the case without kinetic parameter fluctuations. However, when $\sigma_{A}=\sigma_{I}$, the expression reduces to
\begin{align}
   \braket{C_A}_{pf}= \frac{\underline K_{I}}{\underline K_{I}+\underline K_{A}},
\end{align}
this result coincides with the case in which there are no kinetic parameter fluctuations ($\sigma_{\rho} \to 0$). This occurs because fluctuations in the transition from the inactive to active state are compensated for by fluctuations in the inverse transitions. Therefore, if the kinetic parameter fluctuations exhibit this behavior, it becomes very difficult to identify how kinetic parameter fluctuations affect the mean of $C_A$ \cite{Jiao}.

\begin{figure}[h]
\centering
\includegraphics[width=0.85\columnwidth]{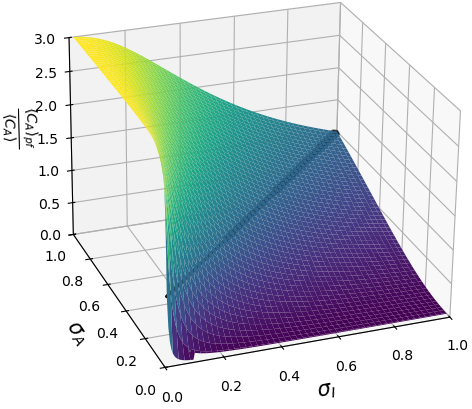}
\caption{ \textbf{Mean active state with gamma distribution.} In this figure, we plot the mean of the active state under kinetic parameter fluctuations that follow the gamma distribution. We observe that these fluctuations can attenuate the mean of the active state or, in the opposite extreme, increase it. We used $\underline K_{I}=1$ and $\underline K_{A}=2$.  }
\label{fig:3}
\end{figure}

In Fig.~\ref{fig:3}, we plot the mean occupancy of the active state as a function of the magnitude of the kinetic parameter fluctuations. We observed that the extrinsic fluctuations can either increase or decrease the mean activity, depending on their magnitude and the transition rate affected.

In particular, increasing $\sigma_A$ increases the mean active-state occupancy. This indicates that fluctuations in the rate of return from the active to the inactive state can increase the time spent in the active state. Conversely, increasing $\sigma_I$ decreases the mean active-state occupancy, indicating that fluctuations in the inactive-to-active transition rate can favor longer periods of inactivity. These results show that extrinsic fluctuations can modify the balance between the active and inactive states without necessarily changing the mean transition rate. However, at a coarse level of observation, different combinations of transition rate fluctuations may produce activity levels similar to those obtained in the absence of such fluctuations.

Consequently, inferring the underlying kinetic parameters from activity and inactivity alone can be challenging, as different combinations of mean transition rates and fluctuation magnitudes may lead to similar stationary distributions \cite{Jiao}.
\section{Parameter Inference and Data Validation} \label{sec:5}

To validate the proposed theoretical model, we conducted an observational study of a domestic cat, treating its observed behavior as a realization of an underlying stochastic process. Data were collected during multiple 60-min observation sessions at different times of the day. Behavioral states were manually annotated through direct observation, following the definitions introduced in Section II. Briefly, the inactive state included resting and sleeping, whereas the active state comprised sustained, observable activities such as walking, grooming, or observing the surroundings. An active episode was considered to begin when the cat stood up and initiated activity, whereas short pauses during an active episode were not classified as transitions to inactivity.

The dataset consists of the following:

\begin{itemize}

\item $\{ T_{K_A}^{(1)}, \dots, T_{K_A}^{(N_A)} \}$: residence times in the active state,

\item $\{ T_{K_I}^{(1)}, \dots, T_{K_I}^{(N_I)} \}$: residence times in the inactive state.

\end{itemize}

\begin{figure}[h]
\centering
\includegraphics[width=0.85\columnwidth]{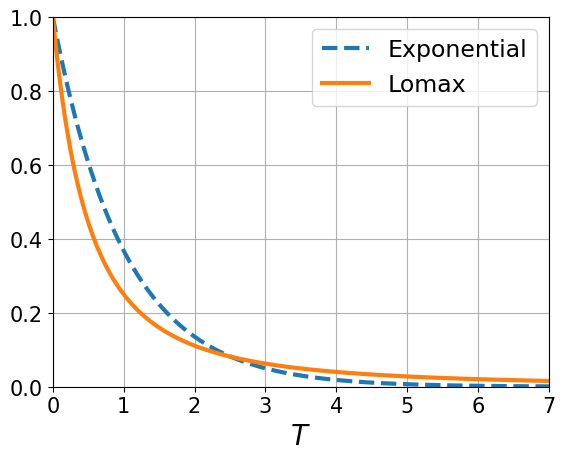}
\caption{ \textbf{Exponential vs Lomax distribution.} In this figure, we plot the exponential distribution \eqref{22} and Lomax distribution \eqref{24}. We observed that the Lomax distribution allows for a higher probability of large values than the exponential distribution, reflecting its heavy-tailed nature. The parameters used are $\underline{\rho}=\rho=1$ and $\sigma_{\rho}=1$.  }
\label{fig:6}
\end{figure}

We assume that $T_{K_A}$ and $T_{K_I}$ are within each state and that transitions follow a memoryless process governed by rates $K_{I}$ and $K_{A}$. Then, the process can be modeled as Markovian. Accordingly, we assume that they follow an exponential distribution \cite{Ossai} ($\rho \in \theta$)
\begin{align}
    P_{\rho}(T_{\rho})= \rho e^{-\rho T_{\rho}}, \label{22}
\end{align}
but we assume that $\rho$ has fluctuations, which follow a gamma distribution, as shown in Eq. \eqref{13}; then the distribution of the observation is
\begin{align}
    P_{\rho,full}(T_{\rho})= &\int_{0}^{\infty} P_{\rho}(\rho) \rho e^{-\rho T_{\rho}} d\rho \nonumber \\
    =& {\underline\rho}{\left( {\sigma_{\rho}}T_{\rho} + 1 \right)^{-\frac{\underline \rho}{\sigma_{\rho}} -1} }, \label{24}
\end{align}
this corresponds to a Lomax distribution \cite{lmoudden}. In Fig. \ref{fig:6}, we compare the exponential and Lomax distributions. We observed that the Lomax distribution assigns a significantly higher probability to large values of $T_{\rho}$ than the exponential distribution, reflecting its heavy-tailed nature \cite{Ham}. 

\subsection{Parameter Inference via Likelihood without Extrinsic Fluctuations }

First, we present how to obtain the kinetic parameter values when there are no extrinsic fluctuations. To determine $K_{\rho}$, from distribution \eqref{22} and using maximum likelihood estimation (MLE) \cite{Chen, Myung}, let the data $\mathcal{D}$ and the parameters $\theta$, then the likelihood function is 
\begin{align}
    L(\mathcal{D}|\theta)= \prod_{i=1} ^{N_\rho} P_\rho(T_{\rho}^{(i)}|\theta),  
\end{align}
where we evaluate the data ($\mathcal{D}$) in the distribution \eqref{13}, then we use the principle of maximum likelihood estimation in $\ln (L(\mathcal{D}|\theta))$, and we get that the values of the kinetic parameters are
\begin{align}
    \hat\rho=  \left( \frac{1}{N_{\rho}} \sum_i^{N_\rho} T^{(i)}_{\rho} \right)^{-1}, \label{26}
\end{align}
from the dataset, we calculate the value of sum. The interval of boundedness is given by 
\begin{align}
    SB(\hat\rho)= \frac{\hat\rho}{\sqrt{N_\rho}}, \label{26.a}
\end{align}
This expression is the Wald interval \cite{Srisura}, and for an interval of boundedness of 95 $\%$ the interval is $\rho \pm 1.96 SB(\rho)$.

\subsection{Parameter Inference via Likelihood with Extrinsic Fluctuations }

Next, we considered the effect of extrinsic fluctuations. To determine $\underline{\rho}$ and $\sigma_{\rho}$ from the distribution \eqref{24} and using MLE \cite{Chen, Myung}. Let the data $\mathcal{D}$ and the hyparameters $\Theta$, then the likelihood funtion is 
\begin{align}
    L(\mathcal{D}|\Theta)= \prod_{i=1}^{N_\rho} P_{\rho, full}(T_\rho^{(i)}|\Theta)
\end{align}
where we evaluate the data in the distribution \eqref{24} and repeat the procedure described in the previous subsection. We use the principle of maximum likelihood estimation to determine the hyperparameters $\underline \rho$ and $\sigma_{\rho}$, it is necessary to solve the following equations, 

{\footnotesize
\begin{subequations}
    \begin{align}
    N_{\rho}= & \left(1 + \frac{N_\rho}{\sum_{i=1}^{N_{\rho}} \ln \left( 1+ \hat\sigma_{\rho} T_\rho^{(i)}  \right)} \right) \sum_{i=1}^{N_{\rho}} \frac{\hat\sigma_{\rho}T_\rho^{(i)}}{1+\hat\sigma_{\rho}T_\rho^{(i)}}, \\
    \underline{\hat\rho}=& \left( \frac{1}{N_{\rho} \hat\sigma_{\rho}} \sum_{i=1}^{N_{\rho}} \ln \left( 1+ \hat\sigma_{\rho} T_\rho^{(i)}  \right) \right)^{-1}.
\end{align}
\end{subequations}}
Because these equations do not admit a closed-form solution, we solved them numerically. In addition, the confidence bounds of the inferred parameters were computed numerically from the Wald interval \cite{Srisura} for a confidence level of 95$\%$.

\subsection{Synthetic and Experimental Data}

We now have all the tools and quantities required to determine the kinetic parameters and hyperparameters. To evaluate whether these quantities can be identified, we first used synthetic data to validate the inference procedure. We used 200 values of $T_{\rho}$ obtained from the Lomax distribution in Eq. \eqref{24}. 

The error between the inferred and the real parameter is calculated as
\begin{align}
    Error= \big| \frac{Inferred-Real}{Real} \big|\times 100 \%.
\end{align}

First, we inferred the parameters using a model without extrinsic fluctuations. The results are presented in Table \ref{tab:1}, where the inferred parameters show substantial deviations from their known values, indicating that neglecting extrinsic fluctuations can lead to biased estimates of the parameters.

We then performed inference using a model that explicitly accounts for extrinsic fluctuations, using the same synthetic dataset to allow a direct comparison between the two approaches. The results are listed in Table \ref{tab:2}. In this case, the inferred kinetic parameters were recovered with substantially smaller errors, while the fluctuation parameters showed different levels of identifiability.

Overall, accounting for extrinsic fluctuations substantially improved the recovery of the parameters used to generate synthetic data compared with the model assuming fixed transition rates. This result demonstrates that neglecting variability in the transition rates can lead to systematic errors in parameter inference and highlights the importance of incorporating extrinsic fluctuations when estimating the underlying kinetic parameters. 

\begin{table}[h]
    \centering
    \caption{\textbf{Parameters inferred from synthetic data without considering extrinsic fluctuations.} The parameters are estimated from synthetic data using Eqs. \eqref{26} and \eqref{26.a}, assuming constant transition rates and neglecting environmental variability.}
    \label{tab:1}
    \begin{tabular}{|c|c|c|c|}
    \hline
        Parameter & Real & Inferred &Error ($\%$)  \\
    \hline
        $K_{I}$ & $100  \times 10^{-2}$  & $74.961  \times 10^{-2}$ & 25.038 \\ 
        $K_{A}$ & $200  \times 10^{-2}$  & $167.030  \times 10^{-2}$ &  16.484\\ 
    \hline
    \end{tabular}
\end{table}

\begin{table}[h]
    \centering
    \caption{\textbf{Parameters inferred from synthetic data considering extrinsic fluctuations.} The parameters were estimated from the synthetic data, explicitly accounting for variability in the transition rates induced by environmental fluctuations. }
    \label{tab:2}
    \begin{tabular}{|c|c|c|c|}
    \hline
        Parameter & Real & Inferred &Error ($\%$)  \\
    \hline
        $\underline{K}_{I}$ & $100  \times 10^{-2}$  & $107.537  \times 10^{-2}$ & 7.537 \\ 
        $\sigma_{I}$ & $25  \times 10^{-2}$  & $33.652  \times 10^{-2}$ & 34.607 \\ 
        $\underline{K}_{A}$ & $200  \times 10^{-2}$  & $191.632  \times 10^{-2}$ &  4.184\\ 
        $\sigma_{A}$ & $25  \times 10^{-2}$  & $24.184  \times 10^{-2}$& 3.2638 \\
        \hline
    \end{tabular}
\end{table}

\begin{table}[h]
    \centering
    \caption{\textbf{Parameters obtained from experimental data without considering extrinsic fluctuations.} We infer the parameters from experimental data collected from a cat using Eqs. \eqref{26} and \eqref{26.a}.}
    \label{tab:3}
    \begin{tabular}{|c|c|}
    \hline
        Parameter & Inferred  \\
    \hline
        $K_{I}$ & $(15.935 \pm 2.149 ) \times 10^{-2} (min^{-1})$   \\ 
        $K_{A}$ & $(13.405 \pm 1.791) \times 10^{-2} (min^{-1})$   \\ 
    \hline
    \end{tabular}
\end{table}

\begin{table}[h]
    \centering
    \caption{\textbf{Parameters obtained from experimental data considering extrinsic fluctuations.} We infer the parameters from experimental data collected, explicitly accounting for variability in the transition rates induced by environmental fluctuations.}
    \label{tab:4}
    \begin{tabular}{|c|c|}
    \hline
        Parameter & Inferred  \\
    \hline
        $\underline{K}_{I}$ & $(31.293 \pm 4.361) \times 10^{-2} (min^{-1})$    \\ 
        $\sigma_{I}$ & $(18.260 \pm 6.045) \times 10^{-2}(min^{-1})$   \\ 
        $\underline{K}_{A}$ & $(16.299 \pm 1.959) \times 10^{-2}(min^{-1})$    \\ 
        $\sigma_{A}$ & $(3.026 \pm 1.930) \times 10^{-2}(min^{-1})$    \\ \hline
    \end{tabular}
\end{table}

We then applied the same inference framework to experimental data obtained from observations of a domestic cat. Parameter estimation was performed under two scenarios: first, assuming fixed transition rates, and second, explicitly accounting for extrinsic fluctuations.

The inferred parameters for the model without extrinsic fluctuations are reported in Table \ref{tab:3}, and the corresponding estimates obtained when accounting for extrinsic fluctuations are shown in Table \ref{tab:4}. The inferred kinetic parameters have comparable magnitudes under both approaches, although accounting for extrinsic fluctuations leads to a different balance between the active and inactive states. The synthetic data analysis showed that neglecting rate fluctuations can lead to biased parameter estimates, motivating the inclusion of extrinsic variability when analyzing experimental data.

For the model without extrinsic fluctuations, the estimated transition rates $K_I$ and $K_A$ were similar in magnitude. Because the stationary probability of the active state is determined by the relative values of these rates, this result indicates a relatively balanced occupancy of the active and inactive states. When extrinsic fluctuations were included, the inferred parameters indicated a greater occupancy of the inactive state. In addition, the inferred fluctuation magnitude was substantially larger for $K_I$ than for $K_A$, with $\sigma_I>\sigma_A$. Thus, the model suggests greater episode-to-episode variability in the effective inactive-to-active transition rate than in the active-to-inactive transition rate.

\section{Conclusion} \label{sec:6}
In this study, we proposed a two-state stochastic model to describe intermittent transitions between the active and inactive states of domestic cat behavior. The model is based on a telegraph process and extends the standard framework by incorporating extrinsic fluctuations into effective transition rates. These fluctuations are represented as episode-to-episode variability in the rates, while the rates are assumed to remain approximately constant during each behavioral episode.

We showed that fluctuations in $K_I$ and $K_A$ can alter the mean occupancy of the active state, either increasing or decreasing it, depending on their magnitude. Interestingly, when the fluctuations in the two transition rates are equal, the mean active-state occupancy coincides with that obtained in the absence of rate fluctuations. Thus, extrinsic fluctuations may remain undetectable from the stationary state occupancy alone, even when they are present in the underlying dynamics of the system.

We further show that fluctuations in the transition rates modify the residence time statistics. While fixed transition rates lead to exponential residence time distributions, stationary variability in the rates gives rise to effective Lomax distributions with heavier tails. This result provides a direct mechanism through which extrinsic variability can generate long inactive or active episodes without requiring non-Markovian dynamics at the individual level.

Parameter inference was evaluated using both synthetic and experimental data. For synthetic data, explicitly accounting for extrinsic fluctuations substantially improved the recovery of the kinetic parameters used to generate the data, whereas neglecting the rate variability produced larger parameter errors. The framework was then applied to the observations of a domestic cat. The experimental analysis yielded a substantially larger inferred fluctuation magnitude for the inactive-to-active transition rate than for the active-to-inactive transition rate, suggesting greater episode-to-episode variability in effective activation dynamics.

Overall, the proposed framework provides an analytically tractable approach for studying stochastic switching systems in which the transition rates exhibit a stationary extrinsic variability. Although illustrated using active–inactive transitions in domestic cat behavior, the formulation can be extended to other telegraph-like systems in which fluctuating transition rates influence burst and residence time statistics, including those of stochastic gene expression, neuronal activity, epidemic processes, and intermittent search dynamics.

\section*{Acknowledgments}

\noindent We would like to express our sincere gratitude to Lombriz for facilitating data collection and for serving as an exemplary subject throughout the behavioral observation period. We would like to thank Manuel Alejandro Mier-Gómez for reading the manuscript. MEHG acknowledges the financial support of SECIHTI through the program "Becas Nacionales 2023"

\section*{Data Availability}

The programs to infer the parameters from synthetic data and experimental data can be found at the following link: https://github.com/Hill137/cat-activity-inactivity

\section*{Declarations}
The authors declare no conflicts of interest regarding the publication of this article. 

%All data generated or analyzed in this study are included in this published article.

%\addtolength{\textheight}{-12cm}   % This command serves to balance the column lengths
                                  % on the last page of the document manually. It shortens
                                  % the textheight of the last page by a suitable amount.
                                  % This command does not take effect until the next page
                                  % so it should come on the page before the last. Make
                                  % sure that you do not shorten the textheight too much.

%%%%%%%%%%%%%%%%%%%%%%%%%%%%%%%%%%%%%%%%%%%%%%%%%%%%%%%%%%%%%%%%%%%%%%%%%%%%%%%%

%%%%%%%%%%%%%%%%%%%%%%%%%%%%%%%%%%%%%%%%%%%%%%%%%%%%%%%%%%%%%%%%%%%%%%%%%%%%%%%%

%%%%%%%%%%%%%%%%%%%%%%%%%%%%%%%%%%%%%%%%%%%%%%%%%%%%%%%%%%%%%%%%%%%%%%%%%%%%%%%%
%\appendix

%\section{Appendix}

%Appendices should appear before the acknowledgment.

%%%%%%%%%%%%%%%%%%%%%%%%%%%%%%%%%%%%%%%%%%%%%%%%%%%%%%%%%%%%%%%%%%%%%%%%%%%%%%%%
%\newpage
\bibliographystyle{ieeetr}
\bibliography{bibliography} 

\end{document}